\documentclass[conf]{new-aiaa}
\usepackage[utf8]{inputenc}
\usepackage{textcomp}

\usepackage{graphicx}
\usepackage{amsmath}
\usepackage[version=4]{mhchem}
\usepackage{siunitx}
\usepackage{longtable,tabularx}
\usepackage{multirow}
\title{
Estimating mean profiles and fluxes in high-speed turbulent boundary layers using inner/outer-layer transformations
}
\author{Asif Manzoor Hasan\footnote{Ph.D. Candidate, Process and Energy Department}}
\affil{
 Delft University of Technology, Leeghwaterstraat 39, 2628 CB, Delft, The Netherlands}
\author{Johan Larsson\footnote{Professor, Department of Mechanical Engineering. Associate Fellow AIAA.}}
\affil{University of Maryland,
College Park, MD 20742, USA
}
\author{Sergio Pirozzoli\footnote{Professor, Dipartimento
 di Ingegneria Meccanica e Aerospaziale}}
\affil{
Sapienza Università di Roma, 
 Via Eudossiana 18,
00184 Roma, Italy
}
\author{Rene Pecnik\footnote{Professor, Process and Energy Department}}
\affil{ 
 Delft University of Technology, Leeghwaterstraat 39, 2628 CB, Delft, The Netherlands}

\begin{document}

\maketitle
\newcommand{\notes}[1]{{\color{red}\begin{itemize}\addtolength{\itemsep}{-2mm} #1 \end{itemize}}}
\newcommand{\note}[1]{{\color{red}#1}}

\begin{abstract}
\end{abstract}

\section*{Nomenclature}

{\renewcommand\arraystretch{1.0}
\noindent\begin{longtable*}{@{}l @{\quad=\quad} l@{} @{}l @{\quad=\quad} l@{}}

$\tau_w$  & wall shear stress&$Re_\tau$ & $\rho_w u_\tau \delta/\mu_w$, friction Reynolds number \\
$\rho$ &    density  &$Re_\tau^*$ & $\bar\rho u_\tau^* \delta/\bar\mu$, semi-local friction Reynolds number \\
$\mu$& dynamic viscosity &$M_\tau$ & $u_\tau/\sqrt{\gamma R T_w}$, friction Mach number\\
$k$& thermal conductivity&$M_\infty$ & $u_\infty/\sqrt{\gamma R T_\infty}$, free-stream Mach number\\
$c_p$& specific heat capacity at constant pressure &$r$ & recovery factor\\
$\gamma$& specific heat capacity ratio &$c_f$ & $2\tau_w/(\rho_\infty u_\infty^2)$, skin-friction coefficient\\
$T$ & temperature&$c_h$ & $q_w/(c_p\rho_\infty u_\infty (T_w-T_r))$, heat-transfer coefficient \\
$R$& specific gas constant&$q_w$ & wall heat flux \\
$u_\tau$ & $\sqrt{\tau_w/\rho_w}$, friction velocity & $y$  & wall-normal coordinate \\
$u_\tau^*$ & $\sqrt{\tau_w/\bar \rho}$, semi-local friction velocity & $y^*$  & $y/\delta_v^*$, semi-local wall-normal coordinate  \\
$\delta_v$   & $\mu_w/(\rho_w u_\tau)$, viscous length scale & $Pr$ & $\mu c_p/ k$, Prandtl number\\
$\delta_v^*$   & $\bar\mu/(\bar\rho u_\tau^*)$, semi-local viscous length scale & $\kappa$ & von K\'arm\'an constant  \\
$\delta$ & (or $\delta_{99}$), boundary layer thickness & $\Pi$ & Coles' wake parameter \\
$\theta$ & momentum thickness & $C$ & log-law intercept\\
$\delta^*$   & displacement thickness& $\mu_t$ & eddy viscosity \\
$U$  & transformed (incompressible) velocity \\
$u$  & untransformed velocity\\
$Re_\theta$ & $\rho_\infty u_\infty \theta/\mu_\infty$, momentum thickness Reynolds number\\
$Re_{\delta^*}$ & $\rho_\infty u_\infty \delta^*/\mu_\infty$, displacement thickness Reynolds number \\
$Re_{\delta_2}$ & $\rho_\infty u_\infty \theta/\mu_w$ \\
\multicolumn{2}{@{}l}{Subscripts}\\
$w$ & wall & $\infty$ & free-stream\\
$e$ & boundary layer edge ($y=\delta$) &$r$ & recovery\\
\multicolumn{2}{@{}l}{Superscripts}\\
$+$ & wall-scaled & $\overline{(\cdot)}$ & Reynolds averaging  \\

\end{longtable*}}

\section{Introduction}

Accurately predicting drag and heat transfer for compressible high-speed flows is of utmost importance for a range of engineering applications. 
This requires the precise knowledge of the entire velocity and temperature profiles. 
A common approach is to use compressible velocity scaling laws (transformation), that inverse transform the velocity profile of an incompressible flow, together with a temperature-velocity relation.
Current methods \citep{huang1993skin, kumar2022modular} typically assume a single velocity scaling law,
neglecting the different scaling characteristics of the inner and outer layers. 
In this Note, we use distinct velocity transformations for these two regions. In the inner layer, we utilize a recently proposed scaling law that appropriately incorporates variable property and intrinsic compressibility effects \citep{hasan2023incorporating}, while the outer layer profile is inverse-transformed with the well-known Van Driest transformation \citep{van1951turbulent}. The result is an analytical expression for the mean shear valid in the entire boundary layer, which combined with the temperature-velocity relationship in \citet{zhang2014generalized}, 
provides predictions of mean velocity and temperature profiles at unprecedented accuracy. Using these profiles, drag and heat transfer is evaluated with an accuracy of +/-4\% and +/-8\%, respectively, for a wide range of compressible turbulent boundary layers up to Mach numbers of 14.

\section{Proposed method}
An incompressible velocity profile is composed of two parts: (1) the law of the wall in the inner layer, and (2) the velocity defect law in the outer layer.
We can model the law of the wall either by
composite velocity profiles \citep{musker1979explicit,chauhan2007composite,nagib2008variations}, or by integrating the mean momentum equation using a suitable eddy viscosity model \citep{van1956turbulent,johnson1985mathematically}. Here, we follow the latter approach and utilize the Johnson-King \citep{johnson1985mathematically}
eddy viscosity model. Likewise, there are several formulations available to represent the defect law \citep{coles1956law,zagarola1998new,fernholz1996incompressible}, of which we use Coles’ law of the wake \citep{coles1956law}.
 
Once the reference incompressible velocity profile is obtained, we inverse transform it using our recently proposed velocity transformation \citep{hasan2023incorporating} for the inner layer, and the Van Driest (VD) transformation \citep{van1951turbulent} for the outer layer. They are combined as follows:
\begin{equation}\label{velproposedT}
    d\bar u^+ = f_3^{-1}   f_2^{-1} f_1^{-1}
    d \bar U^+_{inner} +  f_1^{-1}d \bar U^+_{wake},
\end{equation}
where the factors $f_1$, $f_2$, and $f_3$ constitute the transformation kernel proposed in \citet{hasan2023incorporating} that accounts for both variable property and intrinsic compressibility effects, given as
\begin{equation}\label{hasantr}
     \frac{d\Bar{U}^{+}}{{d \Bar{u}^+}}=   \underbrace{\left({ \frac{1 + \kappa y^* {D}(M_{\tau})} {1 + \kappa {y^*} {D(0)}}}\right)}_{f_3}\underbrace{{\left({1 - \frac{y}{\delta_v^*}\frac{d \delta_v^*}{dy}}\right)}}_{f_2} \underbrace{\sqrt{\frac{\Bar{\rho}}{\rho_w}}}_{f_1} \, ,
\end{equation}
where
\begin{equation}\label{Dc}
D(M_{\tau}) = \left[1 - \mathrm{exp}\left({\frac{-y^*}{A^+ + f(M_\tau)}}\right)\right]^2.
\end{equation}
The value of $A^+$ differs based on the choice of the von K\'arm\'an constant $\kappa$, such that the log-law intercept is reproduced for that $\kappa$ \citep{nagib2008variations}. With $\kappa=0.41$, the value of $A^+=17$ gives a log-law intercept of 5.2 \citep{iyer2019analysis}, whereas, with $\kappa = 0.384$, $A^+=15.22$ gives a log-law intercept of 4.17. The additive term $f(M_\tau)$ accounts for intrinsic compressibility effects. \citet{hasan2023incorporating} proposed $f(M_\tau) = 19.3 M_\tau$, that is independent of the chosen value of $\kappa$.
In Eq.~\eqref{velproposedT}, $d\bar U^+_{inner}$ is modeled using the Johnson-King eddy viscosity model as $ {dy^*}/[{1+\kappa y^* D(0)}]$, which after integration, recovers the incompressible law of the wall, and $d\bar U^+_{wake} = \Pi/\kappa \sin (\pi y/\delta) \, \pi \, d(y/\delta)$ is the derivative of the Coles' wake function \citep{chauhan2007composite}. 

Inserting the expressions for $d\bar U^+_{inner}$, $d\bar U^+_{wake}$ in Eq.~\eqref{velproposedT}, using $d y^* / dy = f_2/\delta_v^*$, $u_\tau^* = u_\tau f_1^{-1}$, and upon rearrangement, we get the dimensional form of the mean velocity gradient as 
\begin{equation}\label{velproposed}
    \frac{d\bar u}{dy} =  \frac{u_\tau^*}{\delta_v^*}\frac{1}{1+\kappa y^* D(M_\tau)}  +  \frac{u_\tau^*}{\delta} \, \frac{\Pi}{\kappa} \, \pi\, \sin \left(\pi \frac{y}{\delta}\right).
\end{equation}
Eq.~\eqref{velproposed} provides several useful insights. Analogous to an incompressible flow, the mean velocity in a compressible flow is controlled by two distinct length scales, $\delta_v^*$ and $\delta$, characteristic of the inner and outer layers, respectively. The two layers are connected by a common velocity scale $u_\tau^*$ (the semi-local friction velocity), leading to a logarithmic law in the overlap region between them. Moreover, in the overlap layer, the denominator of the first term on the right-hand side reduces to $\kappa y$, consistent with Townsend's attached-eddy hypothesis. 
The second term on the right-hand-side is the wake term accounting for mean density variations, where Coles' wake parameter $\Pi$ depends on the Reynolds number, as discussed in the subsection below.  

\subsection{Characterizing low-Reynolds-number effects on the wake parameter}
\begin{figure}
	\centering
	\includegraphics[width=0.9\textwidth]{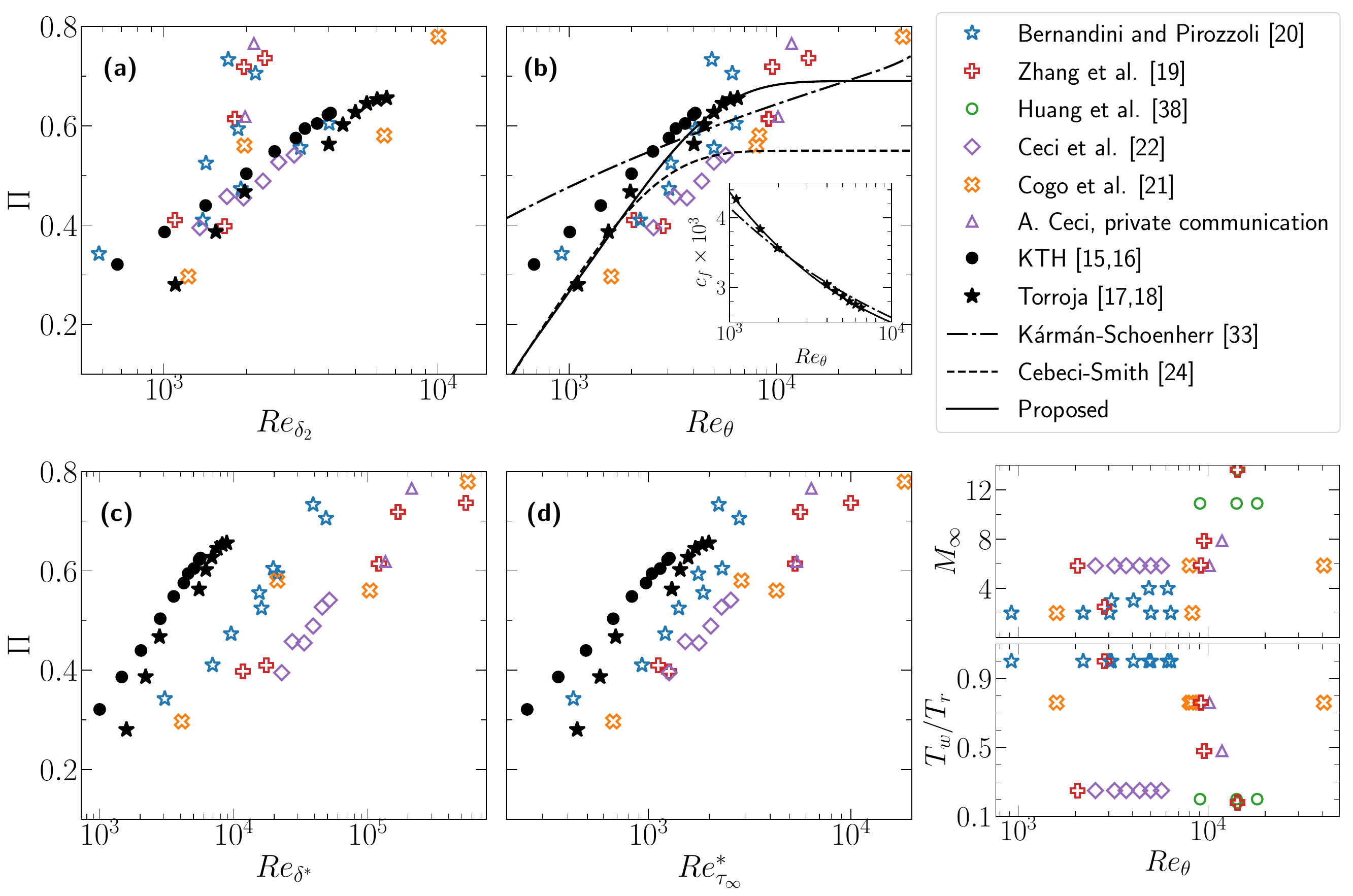}
		\caption{The wake parameter $\Pi$ computed using the DNS data and plotted as a function of (a)~$Re_{\delta_2}$ (b)~$Re_\theta$ (c)~$Re_{\delta^*}$ and (d)~$Re^*_{\tau_\infty}$ for 19 incompressible \citep{schlatter2009turbulent,schlatter2010assessment,jimenez2010turbulent,sillero2013one} and 26 compressible \citep[][A.~Ceci, private communication]{zhang2018direct,bernardini2011wall,cogo2022direct,ceci2022numerical} turbulent boundary layers.}

  \label{fig1}
\end{figure}
For incompressible boundary layers, Coles’ wake parameter is known to strongly depend on $Re_\theta$ at low Reynolds
numbers \citep{coles1962turbulent,fernholz1996incompressible,cebeci2012analysis}. For compressible boundary layers, the ambiguity of the optimal Reynolds number definition poses a challenge to characterize the wake parameter.
\citet{fernholz1980critical}, mainly using experimental data at that time, observed that the momentum-thickness Reynolds number with viscosity at the wall ($Re_{\delta_2}$) is the suitable definition to scale $\Pi$. However, intuitively, $\Pi$ should scale with Reynolds number based on the free-stream properties \citep{smits2006turbulent,cebeci2012analysis}.
Given the recent availability of Direct Numerical Simulation (DNS) data at moderate Reynolds numbers for both compressible and incompressible flows, we revisit the question of which Reynolds number best describes the wake parameter. 

First, we evaluate $\Pi$ for several incompressible and compressible DNS cases from the literature and then report it as a function of different definitions of the Reynolds number, searching for the definition yielding the least spread of the data points. 
For incompressible flows, the wake strength can be determined as ${\Pi} = 0.5{\kappa}\left( \bar U^+(y=\delta) - {1}/{\kappa} \ln(\delta^+) - C\right)$, where $C$ is the log-law intercept for the chosen $\kappa$. For compressible flows, the wake strength is based on the VD transformed velocity \citep{fernholz1980critical,smits2006turbulent} as ${\Pi} = 0.5 \kappa\left( \bar U_{vd}^+(y=\delta) - (\bar U_{vd}^+)^{log}(y=\delta) \right)$,
where $\bar U_{vd}^+$ is obtained from the DNS data. The reference log law $(\bar U_{vd}^+)^{log}$, unlike for incompressible flows, cannot be computed as ${1}/{\kappa} \ln(y^+) + C_{vd}$, because $C_{vd}$ is found to be non-universal for diabatic compressible boundary layers \citep{bradshaw1977compressible,trettel2016mean}. Hence, $(\bar U_{vd}^+)^{log}$ can be obtained either by fitting a logarithmic curve to $\bar U_{vd}^+$ \citep{fernholz1980critical}, or by inverse transforming the incompressible law of the wall.
Here, we follow the latter approach by using the compressibility transformation of \citet{hasan2023incorporating}.

The value of the von K\'arm\'an constant $\kappa$ plays a crucial role in estimating $\Pi$. \citet{spalart1988direct} noted that a strong consensus on $\kappa$ is needed to accurately estimate $\Pi$. However, such a consensus is yet missing \citep{monkewitz2023hunt}. Recently, \citet{nagib2008variations} showed that $\kappa = 0.384$ is a suitable choice for incompressible boundary layers, verified to be true also for channels~\citep{lee2015direct} and pipes~\citep{pirozzoli2021one}. However, due to historical reasons and wide acceptance of $\kappa=0.41$, we will proceed with this value. The same procedure can straightforwardly be repeated with a different value of $\kappa$. 

Figure~\ref{fig1} shows the wake parameter for twenty-six compressible and nineteen incompressible boundary layer flows, as a function of $Re_{\delta_2}$, $Re_{\theta}$, $Re_{\delta^*}$ and $Re_{\tau_\infty}^*$. The spread in the data points is found to be quite large for all the definitions, as $\Pi$ is the difference of two relatively large quantities, namely $\bar U_{vd}^+$ and $(\bar U_{vd}^+)^{log}$ at the boundary layer edge, as outlined above.
Note that even incompressible boundary layers are not devoid of this scatter \citep{spalart1988direct,fernholz1996incompressible}.
Figure \ref{fig1}(a) shows the presence of two distinct branches, hence $Re_{\delta_2}$ does not seem to be suitable to characterize $\Pi$, unlike reported in  previous literature~\citep{fernholz1980critical, huang1993skin}. Among the four definitions of Reynolds number, $Re_\theta$ seems to show the least spread. 
Figure \ref{fig1}(b) also reports several functional forms of $\Pi = f(Re_\theta)$. 
Use of the modified K\'arm\'an-Schoenherr friction formula~\citep{nagib2007approach} for indirect evaluation of $\Pi$, does not show saturation at high Reynolds numbers as observed in \citet{coles1962turbulent} for incompressible flows. The Cebeci-Smith relation \citep{cebeci2012analysis} underpredicts $\Pi$, but reproduces saturation at high Reynolds numbers. We thus propose a relation similar to that proposed by \citep{cebeci2012analysis}, with modified constants to achieve a better fit with data from recent incompressible DNS~\citep{jimenez2010turbulent, sillero2013one}. The relation is
\begin{equation}\label{propfit}
  {\Pi} = 0.69\,\left[1 - \exp(-0.243 \sqrt{z} - {0.15} \,z)\right], \quad\mathrm{where}\quad z = Re_\theta/425 - 1.     
\end{equation}

Inset in Fig.~\ref{fig1}(b) compares the skin-friction curve computed using Eq.~\eqref{propfit} with the modified K\'arm\'an-Schoenherr skin-friction formula \citep{nagib2007approach}.
The distance between the two curves is large at low Reynolds numbers, but less so at higher Reynolds numbers.
As expected, the incompressible DNS data of \citet{jimenez2010turbulent, sillero2013one}, follow the friction curve computed using Eq.~\eqref{propfit}.

\subsection{Implementation of the proposed method}
For convenience, Eq.~\eqref{velproposed} can also be expressed in terms of the dimensional variables $\tau_w$, $\bar\mu$ and $\bar \rho$ as,
\begin{equation}\label{velproposeddim}
    \frac{d\bar u}{dy} =  \frac{\tau_w}{\bar\mu + \underbrace{\sqrt{\tau_w \bar\rho} \kappa y D(M_\tau)}_{\mu_t}}  +  \frac{\sqrt{\tau_w/\bar\rho}}{\delta} \, \frac{\Pi}{\kappa} \, \pi\, \sin \left(\pi \frac{y}{\delta}\right),
\end{equation}
where $\mu_t$ is the Johnson-King eddy viscosity model corrected for intrinsic compressibility effects, derived from  \citet{hasan2023incorporating} transformation. It can be readily used in turbulence modeling, for instance, as a wall-model in Large Eddy Simulations. Note that different eddy viscosity models can be used in Eq.~\eqref{velproposeddim}, for example, Prandtl's mixing length model (see Appendix A). 
Eq.~\eqref{velproposeddim} covers the entire boundary layer, and it can be integrated in conjunction with a suitable temperature model such as the one proposed by \citet{zhang2014generalized}, which is given as
\begin{equation}\label{zhang}
\frac{\bar T}{T_w} =1+\frac{T_r-T_w}{T_w} \left[(1-s\,{Pr})\left(\frac{\bar u}{u_\infty}\right)^2+s \, {Pr}\left(\frac{\bar u}{u_\infty}\right)\right]+\frac{T_\infty-T_r}{T_w}\left(\frac{\bar u}{u_\infty}\right)^2,
\end{equation}
where $s\,{Pr}=0.8$, $ T_r/T_\infty = 1 + 0.5 r (\gamma-1)M_\infty^2$, and $r=Pr^{1/3}$. Moreover, a suitable viscosity law (e.g., power or Sutherland's law), and the ideal gas equation of state $\bar\rho/\rho_w = T_w/\bar T$ have to be used to compute mean viscosity and density profiles, respectively. 
The inputs that need to be provided are the Reynolds number ($Re_\theta$), free-stream Mach number ($M_\infty$), wall cooling/heating parameter ($T_w/T_r$) and (optionally) the dimensional wall or free-stream temperature for Sutherland's law. It is important to note that Eq.~\eqref{zhang}, and all solver inputs are based on the quantities in the free-stream, and not at the boundary layer edge. For more insights on the solver, please refer to the source code available on GitHub~\citep{jupnotebook}.

\section{Results}
\begin{figure}
	\centering	\includegraphics[width=0.9\textwidth]{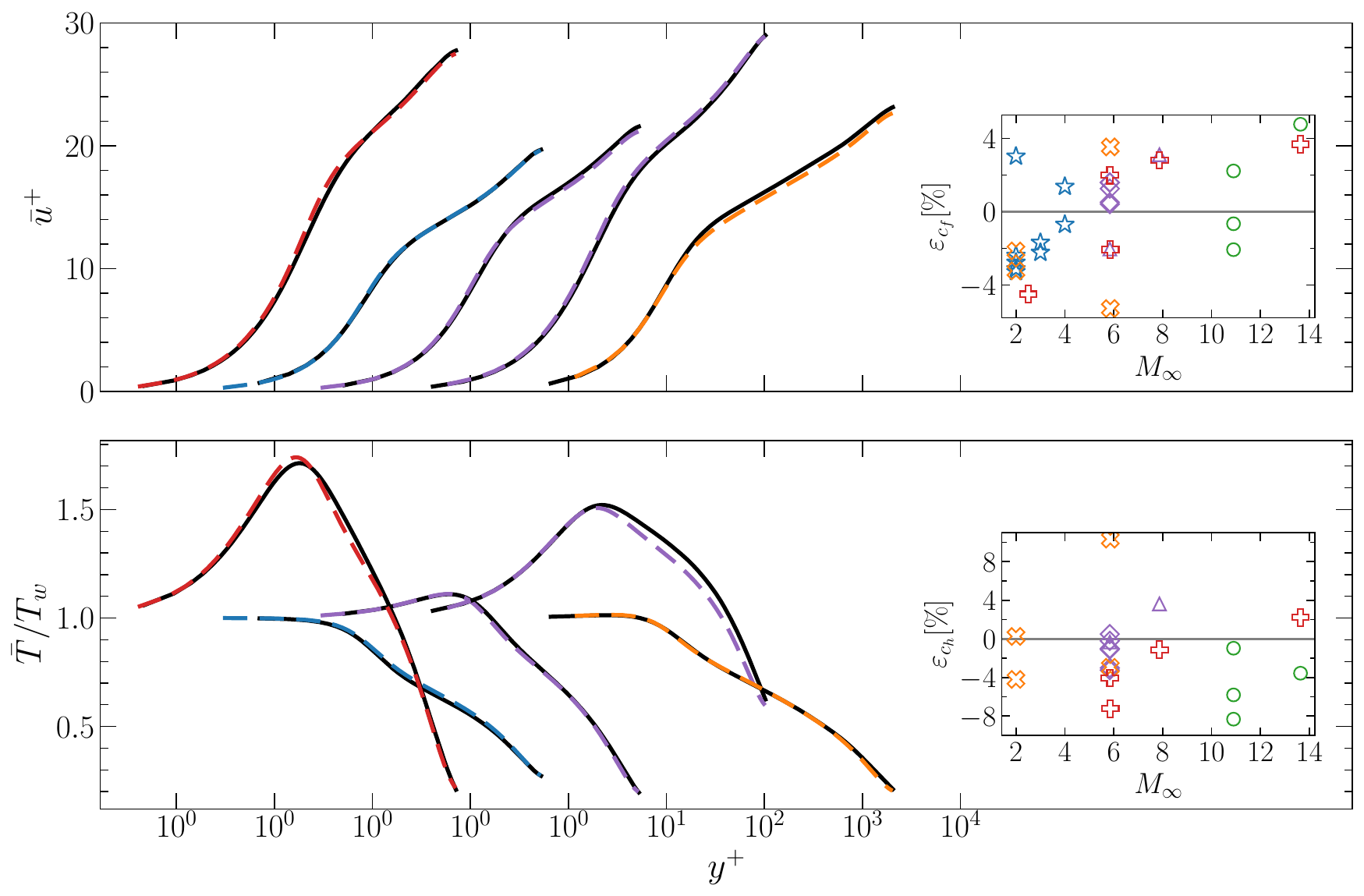}
		\caption{Predicted velocity (top) and temperature (bottom) profiles (dashed lines) compared to DNS (black solid lines) for the cases with the highest reported Mach numbers in the respective publications: (left to right) $M_\infty=13.64$, $T_w/T_r=0.18$ \citep{zhang2018direct}; $M_\infty=4$, $T_w/T_r=1$ \citep{bernardini2011wall}; $M_\infty=7.87$, $T_w/T_r=0.48$ (A.~Ceci, private communication); $M_\infty=5.84$, $T_w/T_r=0.25$ \citep{ceci2022numerical}; $M_\infty=5.86$, $T_w/T_r=0.76$ \citep{cogo2022direct}. (Insets): Percent error in skin-friction (top) and heat-transfer (bottom) prediction for 30 compressible turbulent boundary layers from the literature \citep[][A.~Ceci, private communication]{bernardini2011wall, zhang2018direct, ceci2022numerical, cogo2022direct,huang2020simulation}. The error is computed as $\varepsilon_{c_f} = (c_f - c_f^{DNS})/c_f^{DNS} \,\times 100$ and likewise for $\varepsilon_{c_h}$.
  Symbols are as in Figure~\ref{fig1}. 
  }
  \label{fig2}
\end{figure}

Figure \ref{fig2} shows the predicted velocity and temperature profiles for a selection of high Mach number cases. As can be seen, the DNS and the predicted profiles are in good agreement, thus corroborating our methodology. The insets in Figure \ref{fig2} show the error in the predicted skin-friction and heat-transfer coefficients for thirty compressible cases from the literature. For most cases, the friction coefficient ($c_f$) is predicted with $+/-4\%$ accuracy, with a maximum error of -5.3\%. The prediction of the heat-transfer coefficient ($c_h$) shows a slightly larger error compared to $c_f$, potentially due to additional inaccuracies arising from the temperature-velocity relation. In most cases, $c_h$ is predicted with $+/-8\%$ accuracy, with a maximum error of 10.3\%.

The proposed method is modular in that it can also be applied using other inner-layer transformations \citep{griffin2021velocity, volpiani2020data, trettel2016mean} with minor modifications as discussed in Appendix B. This is shown in Figure \ref{fig3}, which compares the proposed approach with another modular approach of \citet{kumar2022modular}, both with different inner layer transformations. Additionally, the figure includes results obtained with the method of \citet{huang1993skin} using the VD transformation, and the widely recognized Van Driest II skin-friction formula \citep{van1956problem}. 
Figure \ref{fig3} also shows the root-mean-square error, determined as $\mathrm{RMS} = \sqrt{{1}/{N}\sum \varepsilon_{c_f}^2},$
where $N$ is the total number of DNS cases considered. The Van Driest II formula and the method of \citeauthor{huang1993skin} have similar RMS error of about 6\% \footnote{\citeauthor{huang1993skin}'s method with the more accurate temperature velocity relation in \citet{zhang2014generalized} leads to an RMS error of 12\%.}, which is not surprising as both of them are built on Van Driest's mixing-length arguments. The errors are selectively positive for majority of the cases, and it increases with higher Mach number and stronger wall cooling. 
The source of this error mainly resides in the inaccuracy of the VD velocity transformation in the near-wall region for diabatic flows.  
To eliminate this shortcoming, \citet{kumar2022modular} developed a modular methodology, which is quite accurate when the transformation of \citet{volpiani2020data} is used, but it is less accurate if other velocity transformations are implemented. 
This inaccuracy is because the outer layer velocity profile is also inverse-transformed according to the inner-layer transformation.
In the current approach, the velocity profile is instead inverse-transformed using two distinct transformations, which take into account the different scaling properties of the inner and outer layers, thus reducing the RMS error with respect to Kumar and Larsson's modular method for all the transformations tested herein. 
The error using the proposed approach with the TL transformation is preferentially positive for all the cases. This is due to the log-law shift observed in the TL scaling, which is effectively removed in the \citeauthor{hasan2023incorporating} transformation, thereby yielding an RMS error of 2.66\%, which is the lowest among all approaches.

\begin{figure}
	\centering	\includegraphics[width=1\textwidth]{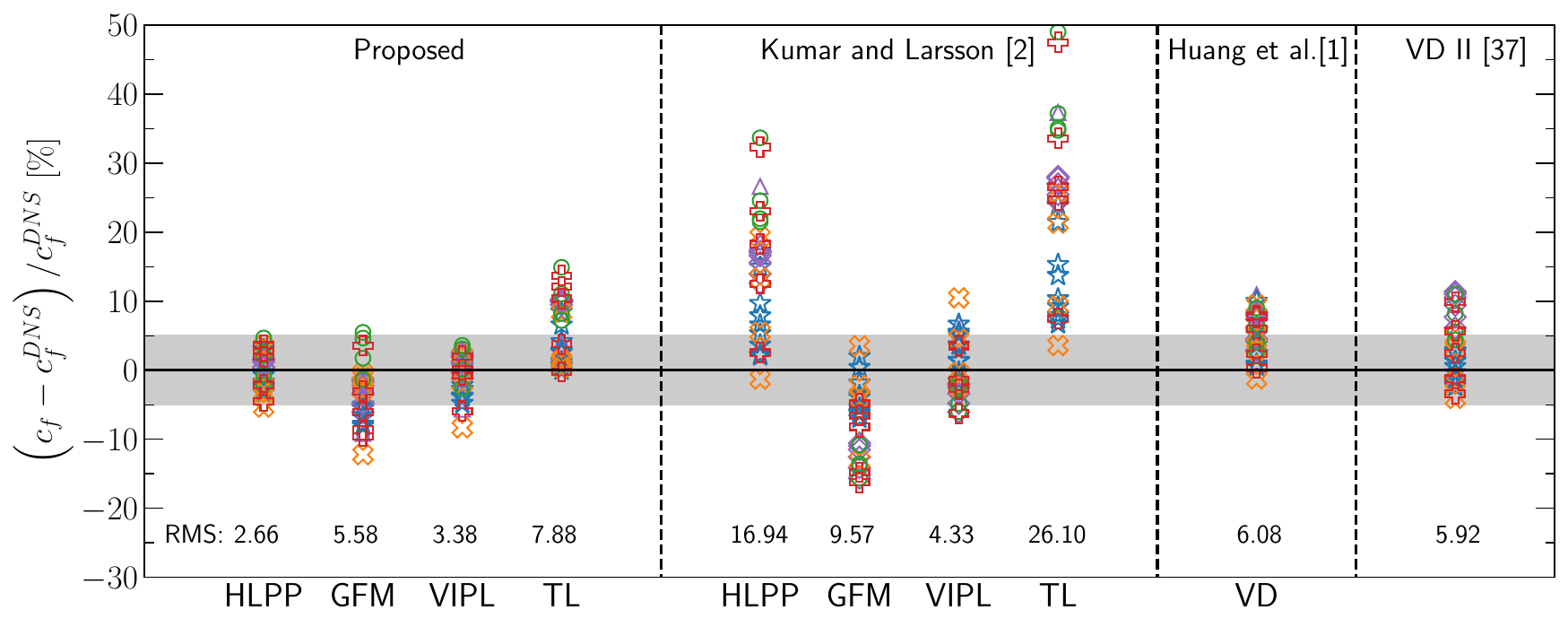}
		\caption{Error in skin-friction prediction using the proposed approach compared to different state-of-the-art approaches. The letters on the X-axis denote the velocity transformation used for that approach. HLPP, GFM, VIPL, TL, and VD stand for the transformations proposed in \citet{hasan2023incorporating}, \citet{griffin2021velocity}, \citet{volpiani2020data}, \citet{trettel2016mean}, and \citet{ van1951turbulent}, respectively.
  The numbers are RMS values computed as outlined in the text. 
  Symbols are as in Figure~\ref{fig1}. 
  The shaded region shows an error bar of +/-5\%. Note that inputs for all the methods are based on properties in the free-stream instead of at the edge of the boundary layer.}
  \label{fig3}
\end{figure}

\section{Conclusions}

We have derived an expression for the mean velocity gradient in high-speed boundary layers [Eq.~\eqref{velproposeddim}] that combines the inner-layer transformation recently proposed by \citet{hasan2023incorporating} and the \citet{van1951turbulent} outer-layer transformation, thus covering the entire boundary layer. The Coles' wake parameter in this expression is determined using an adjusted Cebeci and Smith relation [Eq.~\eqref{propfit}] with the definition of  $Re_\theta$ as the most suitable parameter to characterize low-Reynolds-number effects on $\Pi$. 
This method allows remarkably accurate predictions of the mean velocity and temperature profiles, leading to estimation of the friction and heat-transfer coefficients which are within $+/-4\%$ and $+/-8\%$ error bounds with respect to DNS data, respectively.

The skin-friction results are compared with that from other state-of-the-art approaches considered in literature. Limited accuracy of the VD II and \citeauthor{huang1993skin}'s methods is attributable to inaccuracy of the underlying VD transformation in the near-wall region of diabatic boundary layers, whereas inaccuracy in \citeauthor{kumar2022modular}'s approach is attributable to the unsuitability of the inner layer velocity transformations in the outer layer. By combining different scaling laws in the inner layer with the Van Driest transformation in the outer layer, our method demonstrates improved results for all the inner-layer transformations herein tested, with the lowest RMS error of 2.66\% achieved with \citeauthor{hasan2023incorporating}'s transformation.

The methodology developed in this note promises straightforward application to other classes of wall-bounded flows like channels and pipes, upon change of the temperature-velocity relation \citep[e.g.][]{song2022central}, and using different values of the wake parameter $\Pi$ \citep{nagib2008variations}.
Also, the method is modular in the sense that it can be used with other temperature models and equations of state. 

\section*{Acknowledgements}
We thank Dr.~P.~Costa for the insightful discussions.
This work was supported by the European Research Council grant no. ERC-2019-CoG-864660, Critical; 
and the Air Force Office of Scientific Research under grants FA9550-19-1-0210 and FA9550-19-1-7029.

\section*{Appendix A: Mean shear using Prandtl's mixing length model} \label{AppA}
The choice of the eddy viscosity model affects the first term on the right hand side of Eq.~\eqref{velproposeddim}. By analogy, the mean shear equation using Prandtl's mixing length model is thus as follows,
\begin{equation}\label{velproposeddimpml}
    \frac{d\bar u}{dy} =  \frac{2\,\tau_w}{\bar\mu + \sqrt{\bar\mu^2+[2\sqrt{\tau_w\bar\rho}\kappa y D(M_\tau)]^2}}  +  \frac{\sqrt{\tau_w/\bar\rho}}{\delta} \, \frac{\Pi}{\kappa} \, \pi\, \sin \left(\pi \frac{y}{\delta}\right),
\end{equation}
where $D(M_\tau)$ is the damping function corrected for intrinsic compressibility effects as
\begin{equation}\label{Dcpml}
D(M_\tau) = 1 - \mathrm{exp}\left({\frac{-y^*}{A^+ + 39 M_\tau}}\right),
\end{equation}
with $A^+ = 25.53$ (or $26$) for $\kappa = 0.41$, and where the additive term $39 M_\tau$ is obtained following similar steps as for the Johnson-King model (see Ref.
\citep{hasan2023incorporating}).

\section*{Appendix B: Implementation of the method using velocity transformations in Ref. 
\citep{griffin2021velocity,volpiani2020data}} \label{AppC}
In the logarithmic region and beyond, the first term on the right-hand side of Eq.~\eqref{velproposeddim} reduces to $\sqrt{\tau_w/\bar\rho}/(\kappa y)$, which is the same as Van Driest's original arguments \citep{van1951turbulent}. It is crucial to satisfy this condition, otherwise the logarithmic profile extending to the outer layer would not obey Van Driest's scaling. The transformations of \citet{griffin2021velocity,volpiani2020data} fail to satisfy this property. To address this issue, we enforce Van Driest's scaling in the outer layer by modifying Eq.~\eqref{velproposedT} as follows
\begin{equation}\label{lotw}
\begin{split}
y^+_T \leq 50 :\quad & d\bar u^+  = \mathcal{T}^{-1} d\bar U^+_{inner}+ f_1^{-1}d \bar U^+_{wake},\\
y^+_T > 50 : \quad &  d\bar u^+  = f_1^{-1} d\bar U^+_{inner}  + f_1^{-1}d \bar U^+_{wake}, 
\end{split}
\end{equation}
where $\mathcal{T}$ denotes the inner-layer transformation kernel and $y^+_T$ is the transformed coordinate. The value of $50$ is taken arbitrarily as a start of the logarithmic region.


\end{document}